\documentclass[letterpaper]{elsart}
\usepackage[margin=10pt,font=small,labelfont=bf,format=hang,indention=1.0cm,flushleft]{caption}
\usepackage{graphicx,amssymb,amsmath,times,wrapfig,units,fancyhdr,lineno}
\usepackage{geometry}
\journal{Astroparticle Physics}
\begin{document}
\begin{frontmatter}
\title{The atmospheric charged kaon/pion ratio using seasonal variation methods
}
\author[Minnesota,OSU]{E.~W.~Grashorn\corauthref{cor}}
\corauth[cor]{Corresponding author.}
\ead{grashorn@mps.ohio-state.edu}
\author[IIT,Oxford]{J.~K.~de~Jong}
\author[ANL]{M.~C.~Goodman}
\author[Duluth]{A.~Habig}
\author[Minnesota]{M.~L.~Marshak}
\author[Indiana]{S.~Mufson}
\author[Oxford]{S.~Osprey}
\author[Benedictine]{P.~Schreiner}


\address[ANL]{Argonne National Laboratory, Argonne, Illinois 60439, USA}
\address[Benedictine]{Physics Department, Benedictine University, Lisle, Illinois 60532, USA}
\address[IIT]{Physics Division, Illinois Institute of Technology, Chicago, Illinois 60616, USA}
\address[Indiana]{Indiana University, Bloomington, Indiana 47405, USA}
\address[Minnesota]{University of Minnesota, Minneapolis, Minnesota
55455, USA}
\address[Duluth]{Department of Physics, University of Minnesota -- Duluth, Duluth, Minnesota 55812, USA}
\address[OSU]{Center for Cosmology and Astro-Particle Physics,
 Ohio State University, Columbus, OH 43210, USA}
\address[Oxford]{Department of Physics, University of
Oxford,  Denys Wilkinson Building, Keble Road, Oxford OX1 3RH, United
Kingdom}

\date{\today}

\begin{abstract}
Observed since the 1950's, the seasonal effect on underground muons is a
well studied phenomenon.  The interaction height of incident cosmic rays
changes  as the temperature of the atmosphere changes, which affects the
production height of mesons (mostly pions and kaons).  The decay of these
mesons produces muons that can be detected underground.  The production of
muons is dominated by pion decay, and previous work did not include the
effect of kaons.
In this work, the methods of Barrett and MACRO
are extended to include the effect of kaons.  These efforts give rise to
a new method to measure the atmospheric K/$\pi$ ratio at energies beyond
the reach of current fixed target experiments.  These methods were
applied to data from the MINOS far detector.  A method is developed
for making these measurements at other underground detectors, including
OPERA, Super-K, IceCube, Baksan and the MINOS near detector.

\end{abstract}

\end{frontmatter}
\section{Introduction }\label{sec:intro}
	
When cosmic rays interact in the stratosphere, mesons
are produced in the primary hadronic shower. These mesons either
interact again and produce lower energy hadronic cascades, or decay into
high energy muons which can penetrate to detectors deep underground.  
The temperature of the stratosphere remains nearly constant, only
changing slowly
over longer timescales such as seasons (with the exception of the 
occasional Sudden
Stratospheric Warming events observed during wintertime at high latitudes~\cite{Osprey:2009}).  An increase in temperature of the
stratosphere causes  a decrease in density, reducing the chance of meson
interaction, resulting in a larger fraction decaying to produce muons.
This results in a higher muon rate observed deep
underground~\cite{Barrett:1952,Ambrosio:1997tc,Bouchta:1999kg,Adamson:2009zf,Tilav:2009}. The
effect increases as higher energy muons are sampled, because higher
energy mesons with increased lifetimes (due to time dilation) are involved.  
This effect permits the measurement of the atmospheric charged kaon/pion
production ratio.
The rate of low energy muons at the surface of the earth is also affected by
the temperature because the varying production altitude changes the
chances of the muon decaying before reaching earth, but this effect is not
relevant for detectors deeper than \unit[50]{mwe}\cite{Ambrosio:1997tc} (meters water
equivalent).

\section{ Muon Intensity Underground}
The intensity of muons underground is directly related to the production
of mesons in the stratosphere by hadronic interactions between
cosmic rays and the nuclei of air molecules.  It is assumed that meson
production falls off exponentially as $e^{-X/\Lambda_N}$ where
$\Lambda_N$ is the absorption mean free path of the 
cosmic rays and X is the slant depth of atmospheric material traversed.
It is also assumed that the mesons retain the same direction as their
progenitors, that the cosmic ray sky is isotropic in solid angle at the
top of the atmosphere, and ionization is neglected. These assumptions are particularly
valid for the large energies of the mesons that produce muons seen in
deep underground detectors such as Baksan~\cite{Alexeyev:1979},
Super-K~\cite{Fukuda:2002uc}, IceCube~\cite{Ahrens:2002dv} and MINOS far
detector (MINOS FD) and near detector (MINOS ND)~\cite{MinosNIM}.  In this approximation, $\Lambda_N$ is constant.
Two meson absorption processes will be considered: further hadronic
interactions, $dX/\Lambda_M$, where dX is the amount of atmosphere
traversed,  and $ M \rightarrow \mu \nu_{\mu}$ decay.  The fractional
loss of mesons by decay is given by
\begin{equation}
\frac{m_Mc}{p_r}\frac{dX}{\rho c \tau_0},
\end{equation}
where $\rho =$ air density, $\tau_0 =$ mean M lifetime (at
rest)~\cite{Barrett:1952}, and $p_r$ is the meson rest frame momentum.  M is either a $\pi$ or K meson (charm and
heavier meson production  doesn't become important until $\sim
\unit[10^5]{TeV}$).  For an isothermal, exponentially vanishing
atmosphere, the atmospheric scale height H(T) = RT/Mg.  The density
$\rho$ is then related to X by $\rho = X \cos \theta / H(T)$.  The
critical energy $\epsilon_M$, the energy that separates the atmospheric
interaction and decay regimes, is given by 
\begin{equation}\label{eq:critical}
\epsilon_M = \frac{m_Mc^{2}H(T)}{c\tau_M}.
\end{equation} 
Since most interactions take place in the first few interaction lengths
\cite{Gaisser:1990vg}, and to first order $H(T)~\approx~H_{0}~=~\unit[6.5]{km}$,
 $\epsilon_{\pi} = \unit[0.115]{TeV}$,  $\epsilon_{K} =
\unit[0.850]{TeV}$, the differential meson intensity 
$\mathcal{M}(E,X,\cos \theta)$ can be
written as a function of X~\cite{Barrett:1952,Gaisser:1990vg}: 
\begin{equation}\label{eq:diffpiprod}
\frac{d \mathcal{M}}{dX} = \frac{Z_{N M}}{\Lambda_N}N_0(E)e^{-X/\Lambda_N} - \mathcal{M}(E,X,\cos \theta)
\left[\frac{1}{\Lambda_M} + \frac{\epsilon_M}{EX \cos \theta}\right] 
\end{equation}
for relativistic M, where $N_0(E)$ is the differential M
production spectrum which has the form $E_M^{-(\gamma+1)}$, $\Lambda_N $
is the nucleon interaction length, and $Z_{N M}$ is the
spectrum-weighted inclusive cross section moment.  
This differential equation is straightforward to solve using an
integrating factor  and rewriting
~\cite{Barrett:1952,Gaisser:1990vg}:
\begin{eqnarray}
\label{eq:piprod}
\mathcal{M}(E,X,\theta) & = &
\frac{Z_{N M}}{\Lambda_{N}} N_{0}(E)
e^{-X/\Lambda_M}X^{-\epsilon_M/E \cos \theta}
\int_0^X X'^{\epsilon_M/E\cos \theta} 
e^{-X'/ \Lambda_{M} '}dX'
\nonumber \\
&=& \frac{Z_{N M}}{\Lambda_{N}} N_{0}(E)
e^{-X/\Lambda_M}X
\times  \bigg\{\frac{1}{\epsilon_M/E\cos\theta +1}
-\frac{X/\Lambda'_{M}}{\epsilon_M/E\cos\theta +2} 
\nonumber \\
& & \hspace*{4.5cm} + \frac{1}{2!} \frac{(X/\Lambda'_{M})^2}
{\epsilon_M/E\cos\theta +3}-... \bigg\} , 
\end{eqnarray}
%
%
%
where $1/\Lambda_{M} ' \equiv 1/\Lambda_{N}-1/ \Lambda_M$. 

Now that an expression for the production and propagation of
mesons through the atmosphere has been found, a function describing the
production  of muons must be found.  
Muons are produced from mesons via
the two body decay process $ M \rightarrow \mu \nu$.  The rest frame
momentum for this decay is $p_r = (1-m_{\mu}^2/m_M^2) m_M/2$ (since the
neutrino has negligible mass).  The differential flux per unit
cross section is proportional to the differential flux per energy, which
can be written
\begin{equation}
\label{eq:dnbydE}
\frac{dn}{dE} = \frac{B m_M}{2 p_r P_L},
\end{equation}
where B is the branching ratio and $P_L$ is the momentum of the decaying
particle in the lab frame.   
The muon production spectrum for meson M parents is given by
Gaisser~\cite{Gaisser:1990vg}: 
\begin{equation}\label{eq:muspectrum}
\mathcal{P}_{\mu}(E,X, \cos\theta) =
\sum_{mesons}\int_{E_{min}}^{E_{max}} \frac{dn(E,E')}{dE}
\frac{\epsilon_M}{EX\cos \theta}\mathcal{M}(E',X,\cos \theta)  dE' .
\end{equation}
Inserting Eq.~\ref{eq:dnbydE} into the muon production spectrum
(Eq.~\ref{eq:muspectrum}) gives
\begin{equation}\label{eq:muspectrum2}
\mathcal{P}_{\mu}(E,X, \cos \theta)
=\sum_{mesons}\frac{\epsilon_M}{X\cos\theta(1-r_M)}\int_{E_{\mu}}^{E_{\mu}/r_M}\frac{dE}{E}\frac{\mathcal{M}(E,X,\cos
\theta)}{E} ,
\end{equation}
where $r_M = m_{\mu}^2/m_M^2$.  Muons are sampled 
by detectors at one particular depth, so the production spectrum must be integrated
over the whole atmosphere to find the energy spectrum of interest.
The relevant energy spectrum is written:
\begin{equation}\label{eq:energyspectrum}
\frac{dI_{\mu}}{dE_{\mu}} = \int_0^{\infty}\mathcal{P}_{\mu}(E,X)dX 
\simeq
C_0 \times E^{-(\gamma+1)}\left(\frac{A_{\pi}}{1+1.1E_{\mu}\cos
\theta/\epsilon_{\pi}}+ 0.635~
\frac{A_K}{1+1.1E_{\mu}\cos \theta/\epsilon_{K}}\right) ,
\end{equation}
where $\gamma=1.7$ is the muons spectral index~\cite{Adamson:2007ww},
the branching ratio $B(K\rightarrow \nu_{\mu}\mu)$ = 0.635, and 
$B(\pi\rightarrow \nu_{\mu}\mu) \simeq$ = 1.
The parameters $A_{\pi,K}$ 
are constants involving the amount of inclusive meson
production in the forward fragmentation region, the masses of the mesons
and muons, and the muon spectral index~\cite{Gaisser:1990vg}:
\begin{equation}
A_{\pi(K)} \equiv \frac{Z_{N,\pi(K)}}{(1-r_{\pi(K)})}
\frac{1-(r_{\pi(K)})^{\gamma+1}}{\gamma+1} .
\end{equation}

The integral of the production spectrum can be written in the
form~\cite{Barrett:1952} 
\begin{equation}\label{eq:Iint}
I_{\mu}(E) =
\int_{E_\mathrm{th}}^{\infty}dE_{\mu}\frac{dI_{\mu}}{dE_{\mu}} .
\end{equation}
This is the total number of muons with energy greater than the
minimum required to reach an underground detector. The threshold
surface energy required for a muon to survive to slant depth
\unit[$d(\theta,\phi)$]{(mwe)} increases exponentially as a function of
d and parameters a(E) and b(E)~\cite{Gaisser:1990vg}.  Since a and b depend on
energy, an iterative procedure can be used to find the threshold
energy~\cite{Adamson:2007ww}: 
\begin{equation}\label{eq:Eth}
E_\mathrm{th}=E^{n+1}_\mathrm{th}(\theta,\phi) = \left(E^n+\frac{a}{b}\right)e^{b d(\theta,\phi)}-\frac{a}{b},
\end{equation}
where the energy-dependent parameters \unit[$a = 0.00195+1.09\times
  10^{-4}\ln(E)$]{GeV cm$^2$/g} and \unit[$b = 1.381 \times 10^{-6} +3.96\times
10^{-6}\ln(E)$]{GeV cm$^2$/g}~\cite{Adamson:2007ww}, at column depth $d(\theta,\phi)$.
The threshold energy at the minimum depth of the detectors considered in
Sec.~\ref{sec:intro} are shown in Table~\ref{tab:etable}.
Eq.~\ref{eq:Iint} is approximated~\cite{Barrett:1952} as
\begin{equation}\label{eq:intensity}
I_{\mu} \simeq C_1\times
E_\mathrm{th}^{-\gamma}\left(\frac{1}{\gamma+(\gamma+1)1.1E_\mathrm{th}\cos
\theta/\epsilon_{\pi}}+\frac{0.054}{\gamma+(\gamma+1)1.1E_{\mathrm{th}}\cos
\theta/\epsilon_{K}}\right),
\end{equation}
where 0.635$\cdot A_{K}/A_{\pi}\cdot$r(K/$\pi$)~=~0.054~\cite{Gaisser:1990vg},~\cite{Adamson:2009zf}
and r(K/$\pi$) is the atmospheric K/$\pi$ ratio. 
\begin{table}[!h]
\begin{center}
\caption {Threshold muon energy for current underground detectors. }
\label{tab:etable}
\begin{tabular}{l c c}\\ \hline
\textbf{Detector}  & \textbf{Min. Depth (MWE)} & \textbf{E$_\mathrm{th}$(GeV)} \\\hline \hline
 MINOS ND~\cite{DeJong:2007zz} & 225 & 51  \\
Baksan~\cite{Ambrosio:1997tc} & 850 & 234  \\
IceCube~\cite{Tilav:2009} & 1450 & 466  \\
MINOS FD~\cite{Adamson:2009zf} & 2100 & 730\\
Super-K~\cite{Fukuda:2002uc} & 2700 & 1196 \\
OPERA~\cite{Ambrosio:1997tc} & 3400 &   1833 \\ \hline
\end{tabular}
\end{center}
\end{table}

\section{Temperature Effect on Muon Intensity}\label{sub:tempeff}
\begin{figure}[!h]
 \begin{center}
 \includegraphics[width=0.95\textwidth]{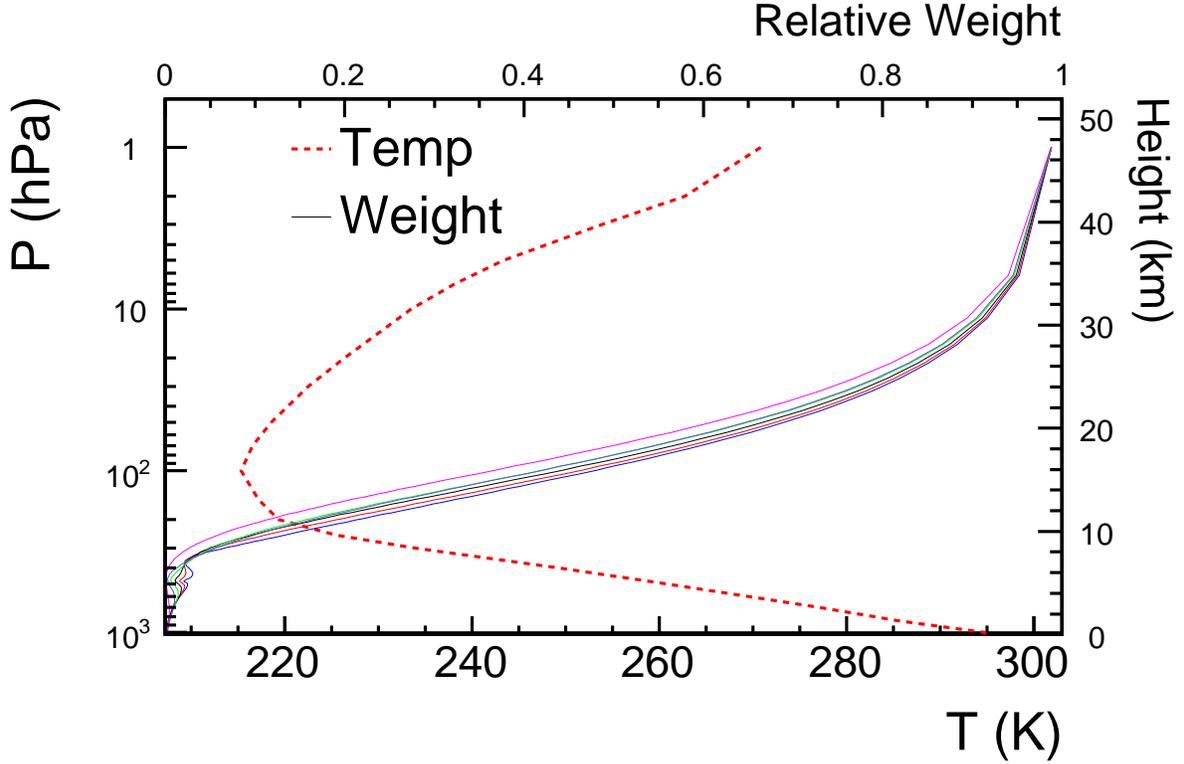}
 \end{center}
 \caption[Temperature Profile]{\label{fig:temp_profile_a}The 
 average mid-latitude summer temperature at various atmospheric depths (dashed
 line). The vertical range is from \unit[1000]{hPa} (\unit[1]{hPa} =
 $\unit[1.019]{g/cm^2}$), near Earth's surface, to \unit[1]{hPa} (nearly
 \unit[50]{km}), near the top of the stratosphere.  The solid lines are 
 the weight as a function of atmospheric depth used to find
 $T_\mathrm{eff}$ (Eq.~\ref{eq:wval}).  
 The blue lines used the OPERA threshold energy
 , the red lines used the Super-K threshold energy
 , the black lines used the MINOS FD threshold energy
 , the green lines used the IceCube threshold energy
 , the violet lines used the Baksan threshold energy 
and the magenta lines used the MINOS ND threshold energy
 .}  
 \end{figure}

\begin{figure}[!h]
 \begin{center}
 \includegraphics[width=0.95\textwidth]{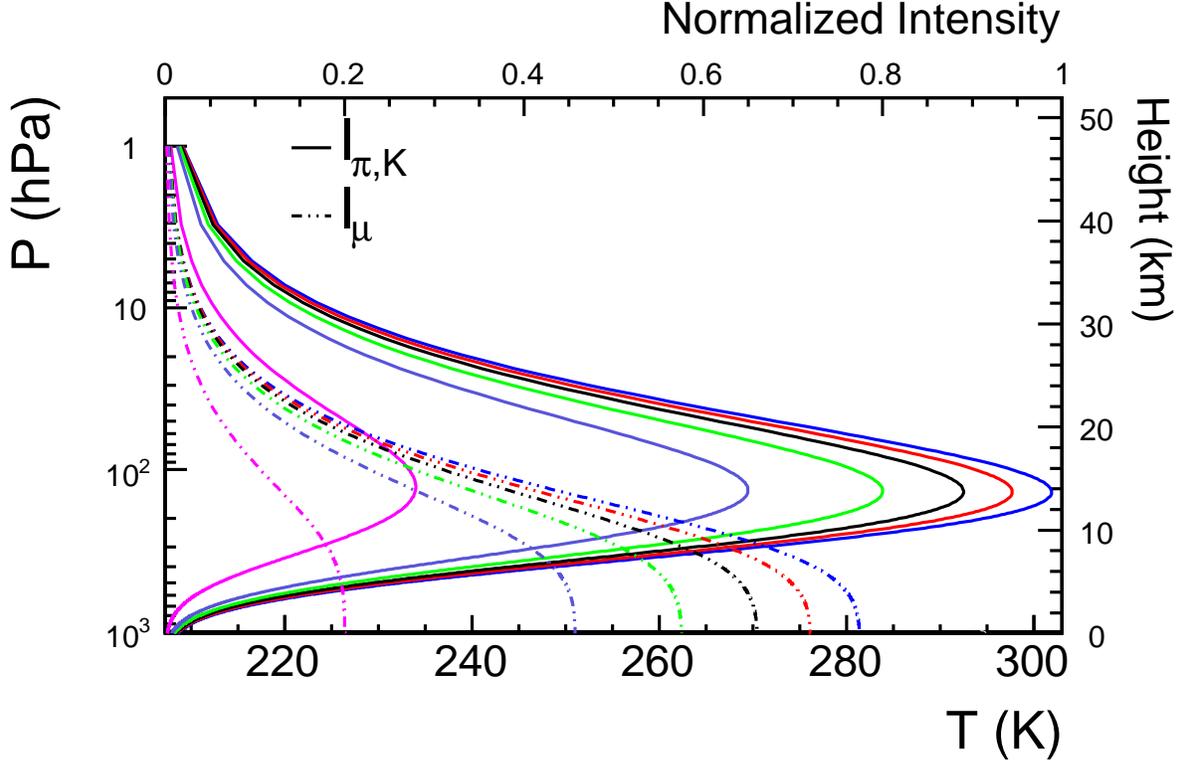}
 \end{center}
 \caption[Temperature Profile]{\label{fig:temp_profile_b}  The solid lines show
 the meson intensity as a function of atmospheric depth
 (Eq.~\ref{eq:piprod}) and the dot-dash lines show the muon intensity as
 a function of atmospheric 
 depth (Eq.~\ref{eq:muspectrum2}).  The normalization in Eq.~\ref{eq:piprod}
 ($N_0 Z_{NM}/\Lambda_N$) was set to 1 to show the dependence on X more
 clearly.  The range of the expressions were adjusted so that the
 maximum value of the OPERA meson intensity was equal to 1, and the
 other equations were scaled appropriately.  These figures were produced
 with particular energy values corresponding to the threshold energy of
 the detectors under consideration.  
 The blue lines used the OPERA threshold energy
 , the red lines used the Super-K threshold energy
 , the black lines used the MINOS FD threshold energy
 , the green lines used the IceCube threshold energy
 , the violet lines used the Baksan threshold energy 
 and the magenta lines used the MINOS ND threshold energy
 .}  
 \end{figure}

The temperature changes that occur in the atmosphere are not
uniform, instead occurring at multiple levels, and neither muon nor
meson production occurs at one particular level (see
Figs.~\ref{fig:temp_profile_a},~\ref{fig:temp_profile_b}).  
The perturbations that variations in temperature cause in
muon intensity are small, 
and as a result properly chosen
atmospheric weights can be used to approximate the effective 
temperature of the atmosphere as a whole, $T_\mathrm{eff}$.   Define
$\eta(X) \equiv (T(X)-T_\mathrm{eff})/T_\mathrm{eff}$, and $\epsilon_M =
\epsilon^0_M(1+\eta)$, where $\epsilon^0_M$ is the constant value of
$\epsilon_M$ when $T=T_\mathrm{eff}$. This is the
temperature that would cause the observed muon intensity if the
atmosphere were isothermal.  To quantify the temperature effect
on intensity, the temperature dependence of Eq.~\ref{eq:critical} needs
to be considered.  The meson production term in
Eq.~\ref{eq:diffpiprod} (which applies to any charged meson: K, $\pi$, etc) can
then be expanded:
\begin{equation}\label{eq:pionspectemp}
\frac{d \mathcal{M}}{dX} = \frac{Z_{N M}}{\Lambda_N}N_0e^{-X/\Lambda_N}
- \mathcal{M}(E,X,\cos \theta) 
\left[\frac{1}{\Lambda_M} + \frac{\epsilon^0_M(1+\eta)} {EX \cos
\theta}\right].  
\end{equation}

The analytic solution to this differential equation is difficult to find
since $\eta(X')$ is an arbitrary function of X'.  A solution to first
order in $\eta(X')$ can be found by expanding the exponential in a power
series, and then following the procedure outlined above, beginning  with
Eq.~\ref{eq:dnbydE}.  This solution can be written as
$\mathcal{M}(E,X,\cos \theta)=\mathcal{M}^0 + \mathcal{M}^1$, where
$\mathcal{M}^0(E,X)$ is the 
solution where $\epsilon_{M} = \epsilon^0_{M}$, which occurs at temperature $T =
T_\mathrm{eff}$ and $\mathcal{M}^1(E,X)$ is given by:
\begin{eqnarray}\label{eq:piprodtemp}
\mathcal{M}^1(E,X,\theta) & = & 
\frac{Z_{N M}}{\Lambda_{N}} N_{0}(E)
e^{-X/\Lambda_M}\left(\frac{X}{\Lambda_M}\right)^{-\epsilon^0_M/E
\cos \theta} 
\frac{\epsilon^0_M}{E \cos \theta}
\int_0^X dX'
\frac{\eta \Lambda_M}{X'}\left(\frac{X'}{\Lambda_M}
\right)^{\epsilon^0_M/E\cos\theta+1}
\nonumber\\
& \times &
\bigg\{\frac{1}{\epsilon^0_M/E\cos\theta +1} 
-\frac{X'/\Lambda'_{M}}{\epsilon^0_M/E\cos\theta +2}
+ \frac{1}{2!}\frac{(X'/\Lambda'_{M})^2}{\epsilon^0_M/E\cos\theta
  +3}-... \bigg\} .  
\end{eqnarray}

If $E \cos \theta \gg \epsilon^0_M$, then the integrand is very
small and $\eta(X') = \eta(X)$.  This is the case when interactions
dominate, as time dilation effects allow these very high energy mesons to
travel great distances before decaying.  If $E \cos \theta \ll
\epsilon^0_M$, then the mesons will not travel very far before decaying and
the integrand is large only when X' is near X, and again, $\eta(X')$ can
be taken out of the integral~\cite{Barrett:1952}. 

Writing the solution of $\mathcal{M}$ where $T = T_\mathrm{eff}$ as
$\mathcal{M}^0$ and letting $\epsilon_M= \epsilon^0_M(1+\eta)$, an
expression for the change in muon  production induced by temperature
variations can be found. Define $\Delta \mathcal{M} \equiv \mathcal{M}-
\mathcal{M}^0$, then to first order in $\eta$
\begin{eqnarray}
\Delta \mathcal{M} =  
\frac{Z_{N M}}{\Lambda_{N}} N_{0}(E)
e^{-X/\Lambda_M}
\frac{\epsilon^0_M\eta X}{E \cos \theta } \times
\bigg\{ & &\frac{1}{(\epsilon^0_M/E\cos\theta +1)^2}
-\frac{2X/\Lambda'_{M}}{(\epsilon^0_M/E\cos\theta +2)^2}
\\ 
&+&\frac{1}{2!}\frac{3(X/\Lambda'_{M})^2}{(\epsilon^0_M/E\cos\theta
+3)^2}-... \bigg\} \nonumber .
\end{eqnarray}
Using Eq.~\ref{eq:muspectrum2} and Eq.~\ref{eq:energyspectrum}, an
expression for the change in differential muon intensity can be found:
\begin{equation}
\Delta \frac{dI_{\mu}}{dE_{\mu}} =
\frac{Z_{N M}}{\Lambda_{N}} 
\left(\frac{\epsilon^0_M} {E_{\mu}\cos\theta }\right)^2
\frac{E_{\mu}^{-(\gamma+1)}}{(1-r_M)}
\int_0^{\infty} dX e^{-X/\Lambda_M}\eta I_M,
\end{equation}
where
\begin{eqnarray}
I_M = 
\int_{1}^{1/r_M}
\frac{dz}{z^{-(\gamma+2)}}  \times & &
\bigg\{ \frac{1}{(\epsilon^0_M/E_{\mu}\cos\theta +z)^2}
-\frac{2X/\Lambda'_{M}}{(\epsilon^0_M/E_{\mu}\cos\theta+2z)^2}
\nonumber\\
& & + \frac{1}{2!}\frac{3(X/\Lambda'_{M})^2}{(\epsilon^0_M/E_{\mu}\cos\theta+3z)^2}-
... \bigg\} .
\end{eqnarray}
Now, a solution to this integral can be found for $E_{\mu} \gg
\epsilon^0_{M} (I^{H})$ and for $E_{\mu} \ll \epsilon^0_{M}(I^{L})$:
\begin{eqnarray}\label{eq:delta_diff_int} 
I^{H}_M(E_{\mu}) &=& 
\frac{1}{\gamma+3}
\left[1-(r_M)^{\gamma+3}\right](1-e^{-X/ \Lambda_{M} '})\frac{\Lambda_{M} '}{X},
\nonumber\\ 
I^{L}_M(E_{\mu}) &=& 
\frac{1}{\gamma+1}
\left[1-(r_M)^{\gamma+1}\right] 
\left(\frac{E_{\mu}\cos \theta}{\epsilon^0_M}\right)^2
(1- X / \Lambda_{M} ')e^{-X/ \Lambda_{M} '} .
\end{eqnarray}

These expressions can be
combined in a form that is valid for all energies 
(Eq.~\ref{eq:energyspectrum}): 
\begin{equation}\label{eq:energy_spectrum_temp}
\Delta \frac{dI_{\mu}}{dE_{\mu}} \simeq \frac{
E_{\mu}^{-(\gamma+1)}}{1-Z_{NN}} \int_0^{\infty}
\frac{ dX (1-X/ \Lambda_{M} ')^2 e^{-X/\Lambda_M}\eta (X) A^1_M}{1+
B^1_M K(X)\left(E_{\mu}\cos\theta/\epsilon^0_M\right)^2} ,
\end{equation}
where
\begin{eqnarray*}\label{eq:AVal}
A^1_K &\equiv& 0.635~\frac{Z_{N,K}}{Z_{N,{\pi}}}
\frac{1-(r_K)^{\gamma+1}}{1-(r_{\pi})^{\gamma+1}} \frac{(1-r_{\pi})}{(1-r_{K})},
\\
A^1_{\pi} &\equiv& 1,
\\
B^1_M &\equiv&
\frac{(\gamma+3)}{(\gamma+1)}\frac{1-(r_M)^{\gamma+1}}{1-(r_M)^{\gamma+3}},
\\
K(X) &\equiv& \frac{(1- X / \Lambda_{M} ')^2}{(1-e^{-X / \Lambda_{M} '})
\Lambda_{M} ' /X} .
\end{eqnarray*}
The exact solution for $ I^{L}_M(E_{\mu})$ has been replaced with an
approximation that preserves the physical behavior of the system at low
energies.   These low energy mesons are relatively insensitive to
changes in temperature because they decay before they have a chance to
interact.  So, this equation describes the expected behavior that mesons
at very low energies  will decay fairly high in the atmosphere.  These 
mesons will not contribute any muons to an
underground detector, because the muons they produce will be below the
threshold energy.

There is a slight dip in this distribution as X approaches $\Lambda'_{M}$,
which results from the approximation made to join the high and low
energy solutions for the approximation to Eq.~\ref{eq:delta_diff_int}.
The low energy solution will go to zero when $X=\Lambda'_{M}$ and below
zero when $E_\mathrm{th} \ll \epsilon^0_{M}$.  The reason for this is that these
low energy muons have such little energy that they decay in flight,
producing a deficit in muons anticorrelated to positive temperature changes (the
``negative temperature coefficient'' related in older
literature~\cite{Barrett:1952,Cini:1967,Humble:1979}).  This effect is
not seen by detectors deeper than \unit[50]{mwe}.  
The fact that there is a dip and subsequent rise in
Fig.~\ref{fig:temp_profile_a} for $X > \unit[480]{g/cm^2}$ does not affect  an
analysis for a detector deeper than \unit[50]{mwe} since the weight is
integrated over the entire atmosphere in discrete steps of dX and
properly normalized, so this atmospheric depth is unimportant for the
production of relevant muons.  

Remembering that $\eta(X) \equiv(T(X)-T_\mathrm{eff})/T_\mathrm{eff}$,
the relationship between atmospheric temperature fluctuations and intensity
variations can be written:
\begin{equation}
\Delta I_{\mu}
= \int_{E_\mathrm{th}}^{\infty} \Delta \frac{dI_{\mu}}{dE_{\mu}} dE_{\mu} 
=\int_0^{\infty} dX \alpha(X) 
\frac{\Delta T(X)}{T_\mathrm{eff}},
\end{equation}
where the temperature coefficient $\alpha(X)$ can be written:
\begin{equation}
\alpha(X) = (1-X/ \Lambda_{M} ')^2 e^{-X/\Lambda_M} \int_{E_\mathrm{th}}^{\infty} dE_{\mu} \frac{A^1_M
E^{-(\gamma+1)}_{\mu}}{1+B^1_M K(X)\left(E_{\mu}\cos\theta / 
\epsilon^0_M\right)^2} = W^M(X) E^{-(\gamma+1)}_\mathrm{th},
\end{equation}
with  $W^M(X)$ given by:
\begin{equation}\label{eq:wval}
W^M(X) \simeq
 \frac{  (1-X/ \Lambda_{M} ')^2  e^{-X/\Lambda_M} A^1_M
}{\gamma +\left( \gamma + 1 \right) B^1_M K(X)\left(E_\mathrm{th}\cos\theta / 
\epsilon^0_M\right)^2} .
\end{equation}
The approximation to the integral follows from arguments made by
Barrett~\cite{Barrett:1952}.  The derivative of this expression agrees
with the integrand in the energy region of interest to within 2\%.
The weights as a function of X  using the threshold energies of the
detectors under consideration can be seen in
Fig.~\ref{fig:temp_profile_a}.  The fact that the lines are nearly on
top of each other shows that the weight of the particular atmospheric
depth does not depend very much on the
threshold energy.  
Recalling that M applies equally to K 
and $\pi$ mesons and that the total muon intensity is the sum of the
contribution by K and $\pi$ (Eq.~\ref{eq:intensity}), the
temperature induced change in muon intensity can be written:
\begin{equation}\label{eq:delta_I_tot}
\Delta I_{\mu}
=\int_0^{\infty} dX \alpha^{\pi}(X) 
\frac{\Delta T(X)}{T_\mathrm{eff}}+
\int_0^{\infty} dX \alpha^{K}(X) 
\frac{\Delta T(X)}{T_\mathrm{eff}} .
\end{equation}

Letting  $T_\mathrm{eff}$ be defined such
that when $T(X)=T_\mathrm{eff}$, $\Delta I_{\mu} = 0$ gives
\begin{equation}\label{eq:teff}
T_\mathrm{eff} = \frac{\int_0^{\infty} dX T(X) \alpha^{\pi}(X)  +
\int_0^{\infty} dX T(X) \alpha^{K}(X) }
{\int_0^{\infty} dX  \alpha^{\pi}(X)  +
\int_0^{\infty} dX  \alpha^{K}(X)} .
\end{equation}
Since the temperature is usually measured at discrete levels, the integral
is calculated numerically over the atmospheric levels $\Delta X_n$:
\begin{equation}\label{eq:teff_exp}
T_\mathrm{eff} \simeq \frac{\sum_{n=0}^{N} \Delta X_n
T(X_n)\left(W_n^{\pi}+W_n^{K}\right)}
{\sum_{n=0}^{N} \Delta X_n \left(W_n^{\pi}+W_n^{K}\right)} .
\end{equation}
$W^{\pi,K}_n $ is $W^{\pi,K} $  evaluated at $X_n$,
$1/\Lambda_{\pi,K} ' \equiv 1/\Lambda_{N}-1/ \Lambda_{\pi,K}$,
$\Lambda_{N} = \unit[120]{g/cm^2}$, $\Lambda_{\pi} = \unit[160]{
g/cm^2}$ and  $\Lambda_{K} = \unit[180]{ g/cm^2}$
\cite{Gaisser:1990vg}. 
%
%

With this definition of Effective Temperature, an Effective
Temperature coefficient, $\alpha_T$, can be defined:
\begin{equation}
\alpha_T = \frac{1}{I^0_{\mu}}\left[ 
\int_0^{\infty} dX  \alpha^{\pi}(X) 
+
\int_0^{\infty} dX \alpha^{K}(X)  \right],
\end{equation} 
where $I^0_{\mu}$ is the intensity for a given temperature T.
Now that the atmospheric temperature has been parametrized and
$\alpha_T$ defined, the relationship between atmospheric temperature
fluctuations and intensity variations can be written:
\begin{equation}\label{eq:alpha1}
\frac{\Delta I_{\mu}}{I^0_{\mu}} = \int_0^{\infty} dX \alpha(X)
\frac{\Delta T(X)}{T_\mathrm{eff}} = \alpha_T \frac{\Delta
T_\mathrm{eff}}{T_\mathrm{eff}} .
\end{equation}
Note that the expression to calculate $T_\mathrm{eff}$ in the pion
scaling limit, ignoring the kaon contribution is the same as the
MACRO~\cite{Ambrosio:1997tc} calculation for effective temperature.  
This distribution reflects the dominant atmospheric phenomena that
produce muons visible to a detector underground.
High energy mesons produced at the top of the atmosphere have the greatest
influence on the seasonal variation
because they are created where the
density is lower, so they have the
highest probability to decay into muons.
High energy mesons that are produced lower in the atmosphere have a
greater probability of interacting a second time, and thus greater
probability of producing muons that are \textsl{not} seen by an
underground detector.

\section{Theoretical Effective Temperature Coefficient}\label{sub:theoretical_alpha}
 The theoretical prediction of $\alpha_T$ for properly weighted
 atmospheric temperature distribution can be written
 (Eq.~\ref{eq:alpha1}):
 \begin{equation}\label{eq:diffalpha}
 \alpha_T = \frac{T}{I^0_{\mu}} \frac{\partial I_{\mu}}{\partial T}.
 \end{equation}
 Barrett~\cite{Barrett:1952} shows that for a muon spectrum such as
Eq.~\ref{eq:energyspectrum}, the theoretical $\alpha_T$ can be written:
\begin{equation}
\alpha_T
 = - \frac{E_\mathrm{th}}{I^0_{\mu}}\frac{\partial I_{\mu}}{\partial E_\mathrm{th}} -
 \gamma .
\end{equation}
The prediction for $\alpha_T$ can be calculated using the intensity found in
Eq.~\ref{eq:intensity} and a little algebra:
\begin{equation}\label{eq:thalpha}
\alpha_T =
\frac{1}{D_{\pi}}
\frac{1/\epsilon^0_{K}+A^1_K(D_{\pi}/D_{K})^2 /\epsilon^0_{\pi} } {
  1/\epsilon^0_{K} + A^1_K(D_{\pi}/D_{K})/\epsilon^0_{\pi}
}, 
\end{equation}
where
\begin{equation}
D_{\pi(K)} = \frac{\gamma}{\gamma+1} \frac{ \epsilon^0_{\pi(K)}}{1.1 E_\mathrm{th} \cos
\theta } + 1 .
\end{equation}
Note that this can be reduced to MACRO's previously
published expression $\left< \alpha_T \right>_{\pi}$
\cite{Ambrosio:1997tc}, which was only valid for pion induced muons, by
setting $A^1_K=0$ (no kaon contribution).  This approximation can be
extended to a kaon-only temperature coefficient, $(\alpha_T)_K$ by
setting the pion term (first term in parentheses) in Eq.~\ref{eq:intensity}.  The result is an
independent model of the temperature coefficient for each of the meson species:
\begin{equation}\label{eq:alpha_th_piExt}
(\alpha_T)_{\pi,K} = 1\big/\left[\frac{\gamma}{\gamma+1} \frac{
\epsilon^0_{\pi,K}}{1.1 E_\mathrm{th} \cos \theta } + 1\right].  
\end{equation}

To compare the experimental $\alpha_T$ to the theoretical
expectation, a simple numerical integration using a Monte Carlo method
was performed. 
A muon energy and $\cos\theta$ were chosen out of the differential
muon intensity~(Eq.~\ref{eq:intensity}).
A random azimuthal angle, $\phi$, was chosen and combined with $\cos
\theta$.  Column depth was calculated as $d = h / \cos\theta$, where h
is the detector depth in mwe for standard rock with flat 
overburden.  The threshold surface energy
required for a muon to survive this column depth is found from
the expression for threshold energy (Eq.~\ref{eq:Eth}).
If the chosen $E_{\mu}$ was greater than $E_\mathrm{th}$, it was used in the
calculation of the theoretical $\left<\alpha_T\right>$.  This was
repeated for 10,000 successful muons with $E_{\mu} > E_\mathrm{th}$, at depths from 0 to \unit[4,000]{mwe}. 
\begin{figure}[h]
\begin{center}
\includegraphics[width = 0.95\textwidth]{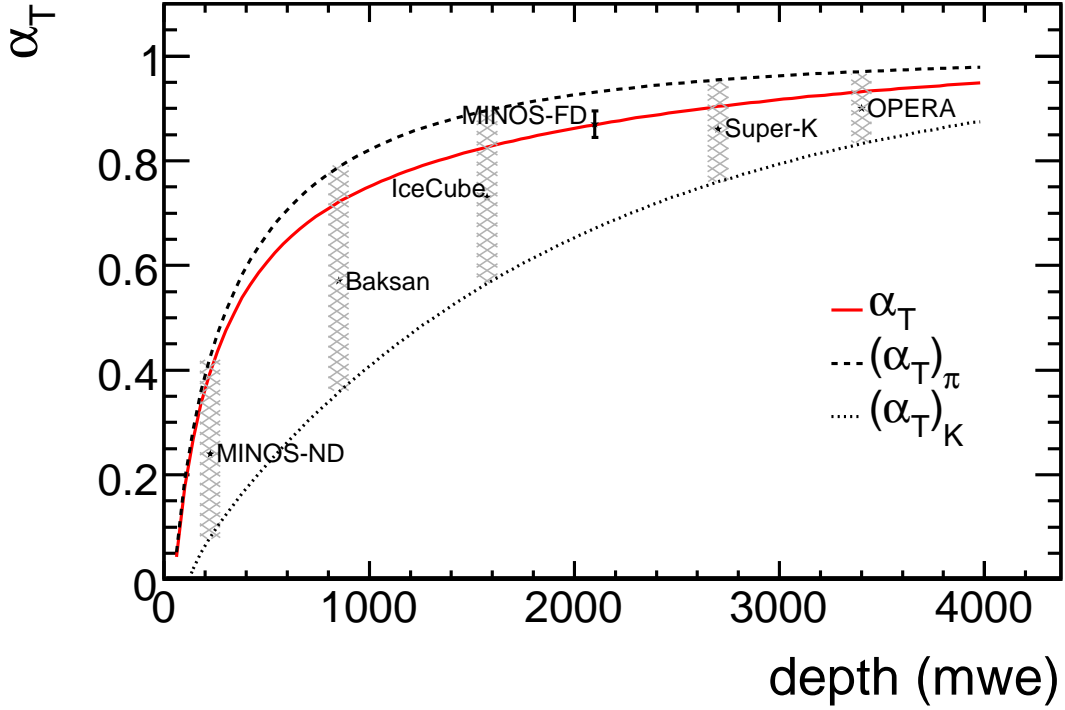}
\vspace{-10pt}
\caption{\label{fig:alphaGlobal}  The theoretical $\left
<\alpha_{T}(X)\right >$ (solid curve), the $\left
<\alpha_{T}(X)\right >_{\pi}$(dashed curve),  the $\left
<\alpha_{T}(X)\right >_K$(dotted curve) for slant depths up to
\unit[4,000]{mwe}.  The MINOS data point is from~\cite{Adamson:2009zf}.
The cross-hatched regions indicate the sensitivity (separation 
between the three models for a particular depth) that
current underground detectors have to measurements of
$\left<\alpha_{T}(X)\right >$.}  
\end{center}
\end{figure}
The result of this calculation, along with the experimental result from
the MINOS experiment~\cite{Adamson:2009zf} can
be seen in Fig.~\ref{fig:alphaGlobal} as the solid line.  This curve
includes the ``negative temperature effect'' (muon decay correction)
term, $\delta ' = (1/E \cos \theta) (m_{\mu} c^2 H / c \tau_{\mu}
)(\gamma / \gamma + 1) \ln (1030/\Lambda_N \cos \theta) $
\cite{Barrett:1952}, which goes to zero for $E_{\mu} > \unit[50]{GeV}$.

The kaon component of air showers that can be observed at
\unit[1400]{mwe} is about 10\%, but the energy is too low for
kaon-induced muon production to be affected by changes in temperature.  The 
result is the large gap between the pion only curve and the K$\pi$
curve.  As the depth increases, the energy of sampled muons also
increases, which results in a greater contribution to $\alpha_T$ by kaon
induced muons. 
The asymptotic behavior of the theoretical $\alpha_T$
approaching one as primary energy increases is expected from
Eq.\ref{eq:diffalpha}.  At very high primary energies, the intensity is
proportional to the critical meson energy,
 which depends on temperature. 
Thus, for an isothermal atmosphere, intensity will be directly
proportional to the temperature (the constant of proportionality,
$\alpha_T$, will be one).

\section{Method for Measurement of Atmospheric K/$\pi$ Ratio}
The theoretical uncertainty of the atmospheric K/$\pi$ ratio is of order
\unit[40]{\%}~\cite{Barr:2006it}.  There was not a measurement of
this ratio with cosmic rays until 2009, when 
it was made using MINOS FD data~\cite{Adamson:2009zf}.  Previous
 measurements have been made at 
accelerators for p+p collisions~\cite{Rossi:1974if}, Au+Au
collisions~\cite{Adler:2002wn}, Pb+Pb
collisions~\cite{Afanasiev:2002mx,Alt:2005zq}. 
 
The expression for the experimental $\alpha_T$ is written in 
Eq.~\ref{eq:alpha1}.
The kaon influence causes an overall decrease in total
$\alpha_T$, shown in Sec.~\ref{sub:theoretical_alpha}. 
Because the left hand side of Eq.~\ref{eq:alpha1} depends only on
counting rate, 
it can be broken into meson components:
\begin{equation}\label{eq:temp_coeff_comp}
 \frac{\Delta R^{\pi}_{\mu}+\Delta R^K_{\mu}}
{\left< R^{\pi}_{\mu} \right> + \left< R^{K}_{\mu}\right>} = \alpha_T\frac{\Delta T_\mathrm{eff}}{\left<T_\mathrm{eff}\right>},
\end{equation}
which can be rewritten:
\begin{equation}
\frac{\left<T_\mathrm{eff}\right>} {\alpha_T \Delta
T_\mathrm{eff}}\left(\frac{\Delta R^{\pi}_{\mu}}
{\left< R^{\pi}_{\mu}\right>}+\frac{\Delta R^K_{\mu}}
 {\left< R^{\pi}_{\mu}\right>}\right)-1 = \frac{
R^{K}_{\mu}}{R^{\pi}_{\mu}} .
\end{equation}

Recall that in the pion scaling limit, only pions are assumed to
contribute to the seasonal effect.  From that, a model for pion-only and
kaon-only temperature coefficients were
developed in Eq.~\ref{eq:alpha_th_piExt}.  Such a 
seasonal effect can be written: 
\begin{equation}\label{eq:pi_temp_coeff2}
 \frac{\Delta R^{\pi,K}_{\mu}} {\left<R^{\pi,K}_{\mu}\right>} =
 (\alpha_T)_{\pi,K} \frac{\Delta T_\mathrm{eff}}
 {\left<T_\mathrm{eff}\right>} . 
\end{equation}
 
The ratio of the muon counting rates $R^K_{\mu}/R^{\pi}_{\mu}$ is
equivalent to the ratio of muons from kaons to muons from pions
$N^K_{\mu}/N^{\pi}_{\mu}$, which will be written $r_{\mu}(K/\pi)$.
Rearranging and inserting Eq.~\ref{eq:pi_temp_coeff2} for both kaons and
pions into Eq.~\ref{eq:temp_coeff_comp} gives:
\begin{eqnarray}
r_{\mu}(K/\pi) &= &\frac{1} {\alpha_T } \left((\alpha_T)_{\pi} +
(\alpha_T)_{K} \frac{\left< R^{K}_{\mu}\right>}{\left<
R^{\pi}_{\mu}\right>} \right) -1
\\& = & \frac{(\alpha_T)_{\pi}/\alpha_T - 1}{1-(\alpha_T)_{K}/\alpha_T} .
\end{eqnarray}
The value for $r_{\mu}(K/\pi)$ can be predicted by integrating
Eq.~\ref{eq:intensity}:
\begin{equation}\label{eq:rmukpi}
 r_{\mu}(K/\pi) = \frac{I_{\mu}^K}{I_{\mu}^{\pi}}
= C_2 \times A^1_K,
\end{equation}
where
\begin{eqnarray}
I_{\mu}^K &=& \int_{E_\mathrm{th} \cos \theta}^{\infty} \frac{
A^1_K~E_{\mu}^{-\gamma}}{1+1.1 E_{\mu} \cos \theta/\epsilon_K}  dE_{\mu} \cos
\theta, \\
I_{\mu}^{\pi} &=& \int_{E_\mathrm{th} \cos \theta}^{\infty} \frac{
E_{\mu}^{-\gamma}}{1+1.1 E_{\mu} \cos \theta/\epsilon^0_{\pi}} dE_{\mu} \cos
\theta.
\end{eqnarray}
The parameter $A^1_K$ is defined as 
\begin{equation}\label{eq:rkpi}
A^1_K = 0.635 \times r(K/\pi) \frac{(1-r_{\pi})}{(1-r_K)}
\frac{1-(r_{K})^{\gamma+1}}{1-(r_{\pi})^{\gamma+1}}, 
\end{equation} 
where $r(K/\pi) = \frac{Z_{NK}}{Z_{N\pi}}$~\cite{Gaisser:1990vg} is the
ratio of kaons to pions produced in the primary cosmic ray 
interactions. 
Inserting Eq.~\ref{eq:rkpi} into Eq.~\ref{eq:rmukpi} and rearranging
gives an expression for $r(K/\pi)$ in terms of $r_{\mu}(K/\pi)$:
\begin{equation} \label{eq:rkpifinal}
r(K/\pi) = \frac{1}{C_1} \times r_{\mu}(K/\pi)\times
\frac{1}{0.635} \frac{(1-r_{K})}{(1-r_{\pi})}
\frac{1-(r_{\pi})^{\gamma+1}}{1-(r_{K})^{\gamma+1}} .
\end{equation}

\begin{figure}[!h]
\begin{center}
\includegraphics[width=.95\textwidth]{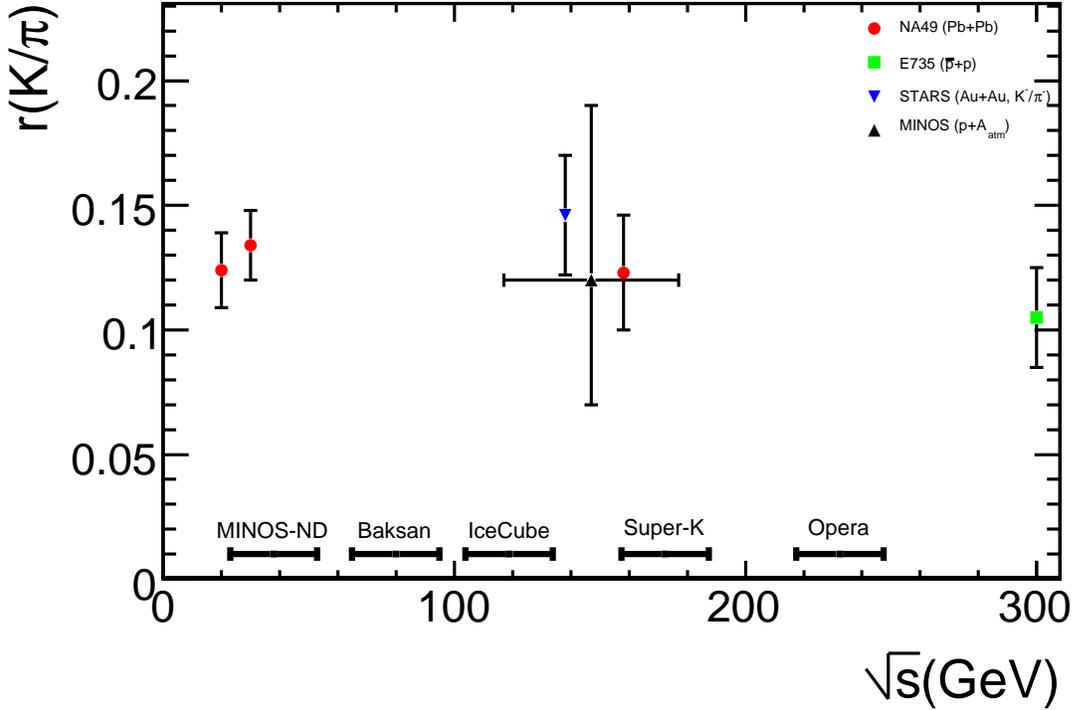}
\end{center}
\caption[ Summary of  K/$\pi$ Measurements]{\label{fig:kpicomp} A
compilation of selected measurements of r(K/$\pi$)  for various primary
particle center of mass energies ($\sqrt{s}$).  The STARS value was from Au+Au collisions at
RHIC~\cite{Adler:2002wn}, the NA49 measurement was from Pb+Pb collisions
at SPS~\cite{Afanasiev:2002mx,Alt:2005zq}, the ISR measurement was
from p+p collisions~\cite{Rossi:1974if}, and the MINOS value was from
cosmic ray primaries + atmospheric nuclei collisions~\cite{Adamson:2009zf}.
The thick horizontal bars near the bottom of the graph
show the typical ranges of cosmic ray primary
energies for the 
collisions that produce muons observed by the underground detectors indicated.
} 
\end{figure}

The MINOS-FD~\cite{Adamson:2009zf} measurement is shown in Fig.~\ref{fig:kpicomp}, along with
STARS~\cite{Adler:2002wn}, NA49~\cite{Afanasiev:2002mx,Alt:2005zq} and
ISR~\cite{Rossi:1974if} accelerator measurements.  The
cross-hatched regions show the energy regimes to which the underground
detectors discussed in this paper are sensitive for $K/\pi$ ratio
measurements using the method described. 
 An OPERA measurement would extend	
to the region an order of magnitude beyond the energy of current fixed target experiments.  
With a detector
area roughly 1,000 times the area of the MINOS FD, IceCube should have a
cosmic ray muon rate of upon completion of construction (Boreal Spring,
2011) of \unit[1,700]{Hz}~\cite{Ahrens:2004ay}.  Assuming a temperature
data set of comparable quality to the BADC ECMWF data~\cite{ECMWF} used by the
MINOS-FD analysis~\cite{Adamson:2009zf}, 
the statistical uncertainty in $\alpha_T$ could be reduced to $\pm
0.001$.  Since the absorber material surrounding IceCube is ice instead
of rock containing iron veins, the column depth should be more well
known.  This could reduce the uncertainty in the depth map, the dominant
source of uncertainty in $(\alpha_T)_{\pi,K}$, by 
half.  These factors taken together could reduce the uncertainty in the
measurement of $r(K/\pi)$ by 16-30\%.  

\section{Summary}
A new method was developed to include the effect of kaons in
measurement of seasonal variations in underground muon intensity.  A
temperature coefficient that accounts for the kaon contribution was
described, and a kaon-inclusive model was defined.  These
methods were applied to MINOS-FD
data~\cite{Adamson:2009zf}, 
and the new model fit the
data better than the pion only model~\cite{Ambrosio:1997tc}.  A formula
was described so that other underground experiments, OPERA, Super-K, IceCube,
Baksan and the MINOS ND could quickly apply this
method to their data. 
Pion and kaon decay are affected differently by temperature variations, and 
this difference suggested a method to measure the atmospheric  K/$\pi$
ratio.  This method was developed and a formula was offered for other
underground experiments to follow.

\section{Acknowledgments}
We thank our many colleagues who provided vital input as these methods
were developed, especially Tom Kelley for providing comments on the
presentation of the mathematics.  This work was supported by the
U.S. Department of Energy,  the U.K. Science and Technologies
Facilities Council, the U.S. National Science Foundation, the Center for
Cosmology and AstroParticle Physics at Ohio State University and the
University of Minnesota.  We also acknowledge the BADC
and the ECMWF for providing the environmental data for this project.
 

\end{document}